\documentclass[aps,prd,showpacs,floatfix,nofootinbib,twocolumn,10pt]{revtex4}

\usepackage{color,amsmath,amssymb,graphicx,latexsym,subfigure}
\usepackage{threeparttable,txfonts}

\begin{document}

\voffset 1.25cm

\title{Understanding the spectral hardenings and radial distribution
of Galactic cosmic rays and Fermi diffuse $\gamma$-rays with
spatially-dependent propagation}

\author{Yi-Qing Guo$^{a}$}
\author{Qiang Yuan$^{b,c}$\footnote{yuanq@pmo.ac.cn}}

\affiliation{
$^a$Key Laboratory of Particle Astrophysics, Institute of High Energy
Physics, Chinese Academy of Sciences, Beijing 100049, P.R.China \\
$^b$Key Laboratory of Dark Matter and Space Astronomy, Purple Mountain 
Observatory, Chinese Academy of Sciences, Nanjing 210008, P.R.China \\
$^c$School of Astronomy and Space Science, University of Science and 
Technology of China, Hefei, Anhui 230026, P.R.China
}

\begin{abstract}

Recent direct measurements of Galactic cosmic ray spectra by
balloon/space-borne detectors reveal spectral hardenings of all major
nucleus species at rigidities of a few hundred GV. The all-sky diffuse
$\gamma$-ray emissions measured by the Fermi Large Area Telescope also
show spatial variations of the intensities and spectral indices of
cosmic rays. These new observations challenge the traditional simple
acceleration and/or propagation scenario of Galactic cosmic rays. In
this work we propose a spatially-dependent diffusion scenario to
explain all these phenomena. The diffusion coefficient is assumed to
be anti-correlated with the source distribution, which is a natural
expectation from the charged particle transportation in turbulent
magnetic field. The spatially-dependent diffusion model also gives
a lower level of anisotropies of cosmic rays, which are consistent
with observations by underground muons and air shower experiments. 
The spectral variations of cosmic rays across the Galaxy can be
properly reproduced by this model.

\end{abstract}

\date{\today}


\maketitle

\section{Introduction}

It has been widely believed that cosmic rays (CRs) below the ``knee''
($\sim$PeV) are originated from Galactic accelerators such as remnants
of supernova explosion \cite{1964ocr..book.....G}. Charged CRs propagate
diffusively in the Milky Way, interact with the interstellar medium (ISM)
and produce secondary particles. Such a ``standard'' paradigm of the
production, propagation, and interaction of Galactic CRs work successfully
to explain most of the observations of CRs as well as diffuse $\gamma$-rays
\cite{2007ARNPS..57..285S}. 

Some recent observations challenge this ``standard'' picture. Remarkable 
spectral hardenings of CR nuclei at several hundred GV have been found by
balloon and space detectors \cite{2007BRASP..71..494P,2010ApJ...714L..89A,
2011Sci...332...69A,2015PhRvL.114q1103A,2015PhRvL.115u1101A}. Several kinds
of models incorporating modifications of simple assumption of the injection, 
acceleration, and propagation of CRs have been proposed to explain it 
(e.g., \cite{2011ApJ...729L..13O,2011PhRvD..84d3002Y,
2012ApJ...752...68V,2012APh....35..449E,2012PhRvL.109f1101B,
2012ApJ...752L..13T,2012MNRAS.421.1209T,2013ApJ...763...47P,
2013A&A...555A..48B,2014A&A...567A..33T,2015ApJ...803L..15T,
2017PhRvD..95b3001T,2017PhRvD..96b3006L,2017ApJ...844L...3Z,2017PhRvD..96d3012D}). 
In addition, the diffuse $\gamma$-ray emission detected by the Fermi Large 
Area Telescope (Fermi-LAT) reveal a flatter CR density gradient toward the 
outer Galaxy region \cite{2011ApJ...726...81A}. The gradient problem might 
imply either a thicker propagation halo of CRs or that there are more sources 
in the outer Galaxy than that inferred from observations of supernova remnants
(SNRs) or pulsars \cite{2011ApJ...726...81A}. Most recently, the analysis 
of the Fermi-LAT diffuse $\gamma$-ray further suggests spatial variations 
of both intensities and spectra of CRs \cite{2016PhRvD..93l3007Y,
2016ApJS..223...26A}, which can not be simply reproduced from the 
conventional CR propagation model\footnote{In this work the conventional
propagation model means the model with uniform, single power-law form 
of the diffusion coefficient, and single power-law source injection 
spectrum above $\sim10$ GV.}. 

It was shown that a spatially-dependent propagation (SDP) scenario can 
account for both the CR intensity gradient and the small anisotropies 
of CR arrival directions \cite{2012PhRvL.108u1102E} (see also
the original work of Ref. \cite{2002JPhG...28.2329E}). 
In Ref. \cite{2016MNRAS.462L..88R}, Recchia et al. proposed a model of
non-linear CR propagation with particle scattering and advection off
self-generated turbulence to account for the spatial variations of the
CR densities and spectra. In this model, the transportation (diffusion
and advection) of CRs varies in the Galaxy amounting to a type of SDP model. 
However, only the one-dimensional diffusion ($z$-direction) of particles 
is assumed \cite{2016MNRAS.462L..88R}. Furthermore, to account for the 
spatial variations of the CR intensities and spectra, an exponential 
decay of the background magnetic field is required. Just recently, 
Cerri et al. proposed an anisotropic diffusion model to interpret 
the radial denpendence of spectra of CRs and suggested that the harder 
slope in the inner Galaxy was due to the parallel diffusive escape along 
the poloidal component of the large-scale, regular, magnetic field 
\cite{2017arXiv170707694C}.

Although there were quite a few studies on using the SDP models to 
understand the newly available CR data \cite{2012ApJ...752L..13T,
2015PhRvD..92h1301T,2016ChPhC..40a5101J,2016ApJ...819...54G,
2016PhRvD..94l3007F,2017arXiv170107136G}, those previous works lack 
a coherent explanation of all the above mentioned observations 
simultaneously. In this work we employ an SDP model of CR propagation 
to self-consistently account for those observations. The diffusion 
coefficient is assumed to be anti-correlated with the CR source 
distribution, which is a natural assumption since the (turbulent) 
magnetic field strength is expected to be correlated with the matter 
distribution. We will show that such a simple extension of the 
conventional CR propagation model can give reasonable fits to most 
of the available data of CRs and $\gamma$-rays.

\section{Model description}

The propagation of charged particles in the Milky Way is usually restricted 
in a cylinder, with a half-height of $z_h$, centered at the Galactic center. 
CRs may further experience convective transportation, reacceleration
due to interactions with random magneto-hydrodynamic waves, energy loss
due to ionization and Coulomb scattering, and/or fragmentation due to 
collisions with the ISM. Secondary nuclei are produced via the
fragmentations of primary nuclei during their propagation. Here we adopt 
the diffusion reacceleration model, which is found to well describe the 
secondary-to-primary ratios and low energy fluxes of CR nuclei 
\cite{2011ApJ...729..106T,2016ApJ...824...16J,2017PhRvD..95h3007Y}, 
to characterize the propagation process of CRs. 

The source distribution of CRs is assumed to follow the observed spatial 
distribution of SNRs
\begin{equation}
f(r,z)=\left(\frac{r}{r_\odot}\right)^{\alpha}\exp\left[-\frac
{\beta(r-r_\odot)}{r_\odot}\right]\,\exp\left(-\frac{|z|}{z_s}\right),
\end{equation}
where $r_\odot=8.5$ kpc, $z_s=0.2$ kpc, $\alpha=1.09$, and $\beta=3.87$ 
\cite{1998ApJ...504..761C}. We have normalized $f(r,z)$ to 1 at the solar 
location. The source spectrum of CRs is assumed to be a broken power-law 
in rigidity.

\begin{figure*}[!htb]
\includegraphics[width=0.45\textwidth]{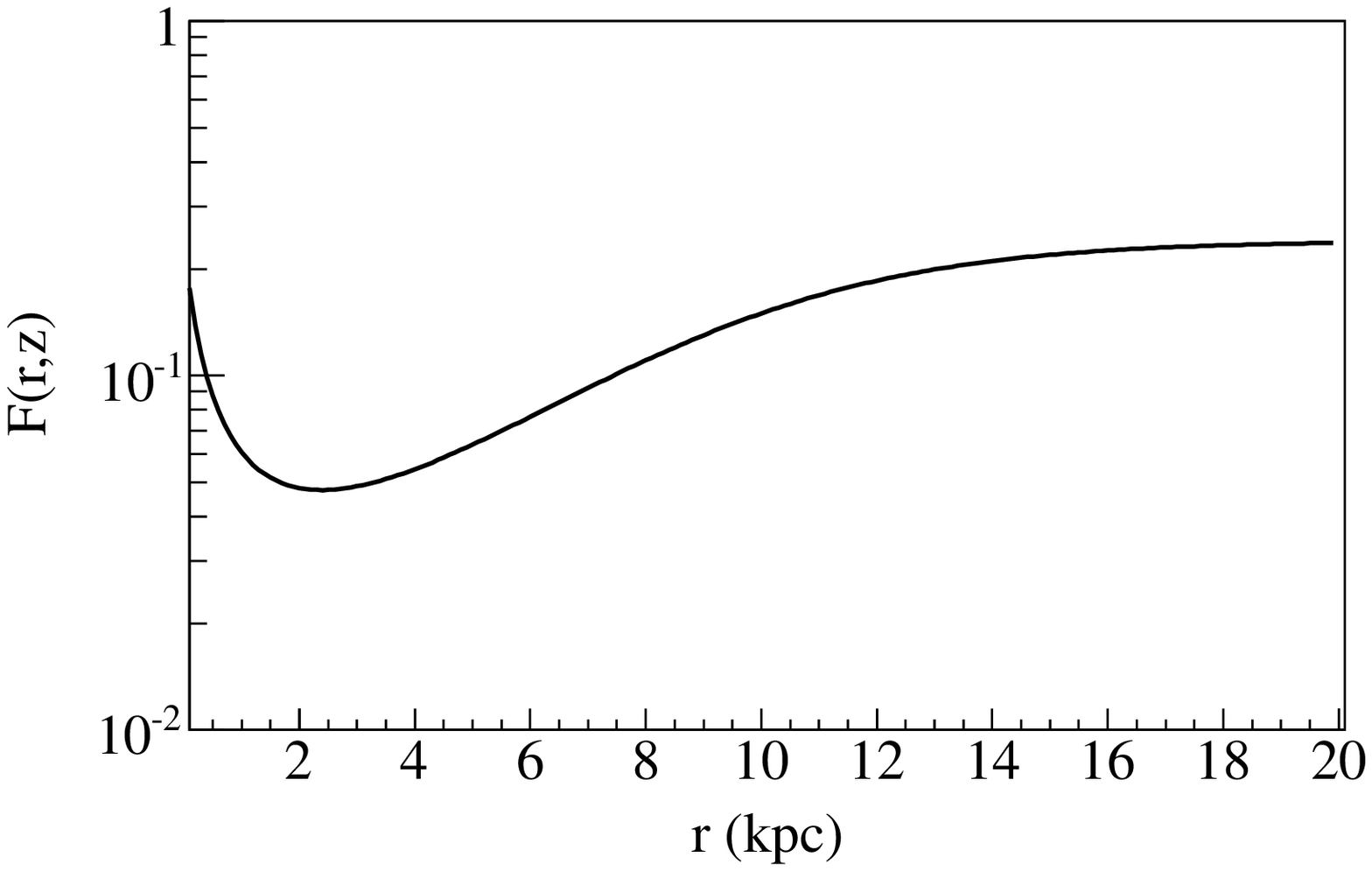}
\includegraphics[width=0.45\textwidth]{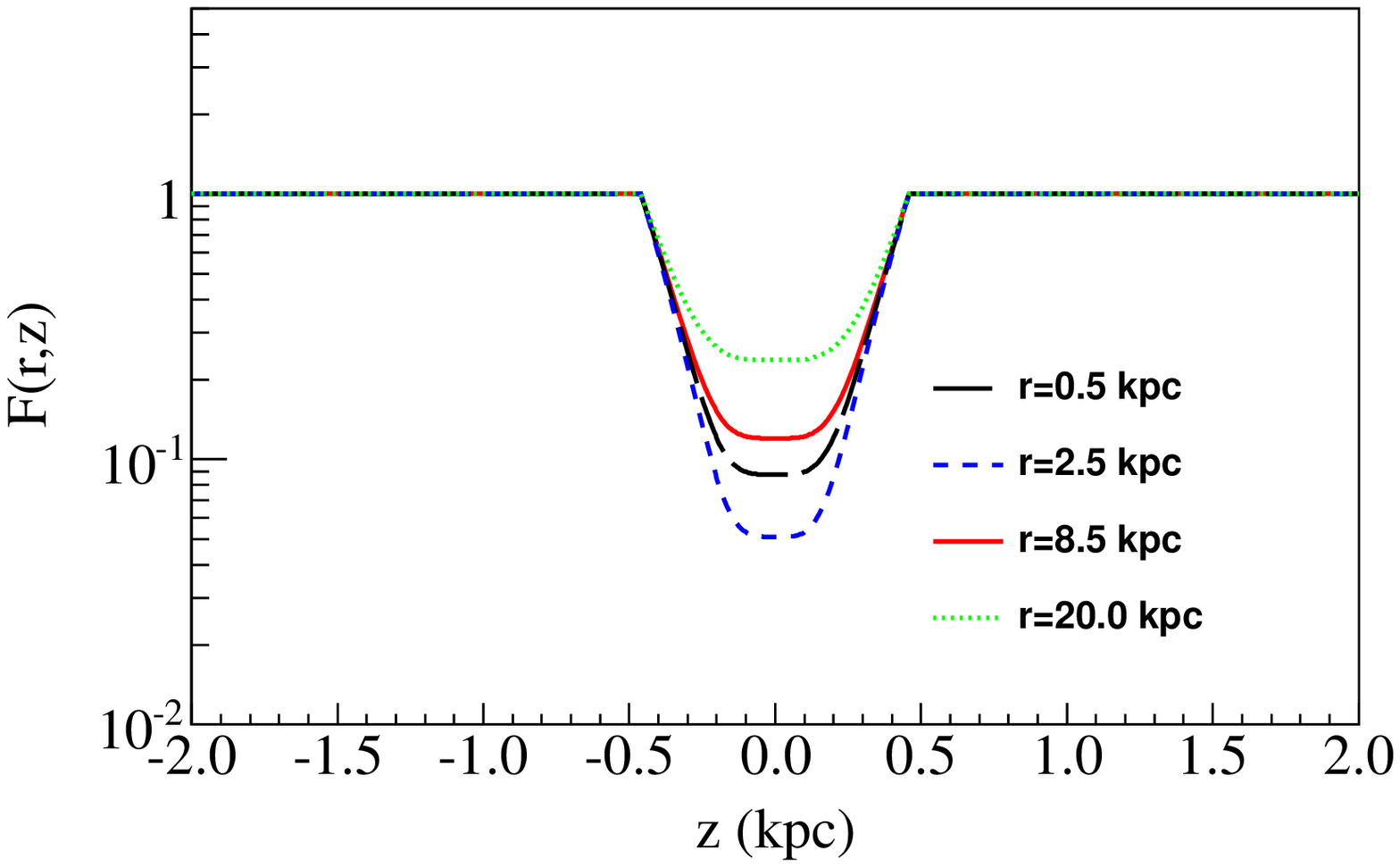}
\caption{The distributions of $F(r,z)$ with the radial distance $r$ 
(left panel; for $z=0$) and vertical height $z$ (right panel).}
\label{fig:SPDparam}
\end{figure*}

The spatial diffusion coefficient is described with a two halo approach: 
the inner (disk) and outer halo \cite{2012ApJ...752L..13T}. The diffusion
coefficient $D_{xx}$ is parameterized as
\begin{equation}
D_{xx}(r,z,p) = F(r,z) D_0 \beta \left(\frac{p}{p_0}\right)^
{F(r,z)\delta_0}, 
\end{equation}
where $F(r,z)D_0$ represents the normalization factor of the diffusion
coefficient at the reference rigidity $p_0$, and $F(r,z)\delta_0$ 
reflects the property of the irregular turbulence. The function $F(r,z)$ 
takes the form as
\begin{eqnarray}
F(r,z)&=&\frac{N_m}{1+f(r,z)} \nonumber\\
      &+& \left(1-\frac{N_m}{1+f(r,z)}\right)\cdot 
      \min\left[\left ( \frac{z}{\xi z_h}\right )^n,1\right],
\end{eqnarray}
where $\xi z_h$ denotes the half thickness of the inner halo, $(1-\xi)z_h$ 
is the half thickness of the outer halo, $N_m$ is a normalization factor, 
and $n$ characterizes the sharpness between the inner and outer halos. 
For $z\ll \xi z_h$ (the inner halo), the diffusion coefficient is obviously 
anti-correlated with the source distribution $f(r,z)$. For the outer halo 
where the source term vanishes, the diffusion coefficient recovers the 
traditional form of $D_0\beta(p/p_0)^{\delta_0}$. To clearly see the 
behaviors of $F(r,z)$, we show their distibutions as functions of $r$ 
(for $z=0$) and $z$ (for a few values of $r$) in Fig. \ref{fig:SPDparam}.

The reacceleration can be characterized by a diffusion in momentum space. 
The momentum diffusion coefficient $D_{pp}$ relates to $D_{xx}$ via the 
effective Alfvenic velocity $v_A$ of the ISM \cite{1994ApJ...431..705S},
as $D_{pp}D_{xx}=\frac{4p^2v_A^2}{3\delta(4-\delta^2)(4-\delta)}$,
where $\delta=F(r,z)\delta_0$.

A numerical method is necessary to solve the diffusion equations,
especially in case that the diffusion varies everywhere in the Milky 
Way. In this work we use the DRAGON code \cite{2008JCAP...10..018E,
2017JCAP...02..015E} to calculate the propagation of CRs. The basic
model parameters are given in Table \ref{table:prop}.

\begin{table}[h]
\begin{center}
\caption{Propagation parameters}
\begin{tabular}{ccccc}
\hline\hline
   $D_0$ (cm$^2$ s$^{-1}$) & $5.6\times 10^{28}$ \\
   $\delta_0$              & 0.56                \\
   $v_A$ (km s$^{-1}$)     & 6.0                 \\
   $z_h$ (kpc)             & 5.0                 \\
   $N_m$                   & 0.24                \\
   $\xi$                   & 0.092               \\
   $n$                     & 5                   \\
\hline \hline
\end{tabular}
\label{table:prop}
\end{center}
\end{table}

\section{Results}

\subsection{Primary CRs}

\begin{figure}[!htb]
\includegraphics[width=0.45\textwidth]{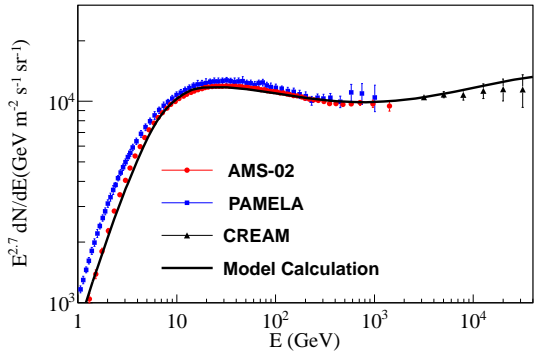}
\caption{Model predictions of the proton spectrum, compared with the
measurements by PAMELA \cite{2011Sci...332...69A}, AMS-02 
\cite{2015PhRvL.114q1103A}, and CREAM \cite{2017ApJ...839....5Y}.
\label{fig:pflux}}
\end{figure}

We first look at the effect on the spectra of primary CRs. Fig. 
\ref{fig:pflux} shows the proton spectrum expected from the SDP model 
and the comparison with the measurements \cite{2011Sci...332...69A,
2015PhRvL.114q1103A,2017ApJ...839....5Y}. It can be seen that the model gives 
a clear hardening of the spectrum for $E\gtrsim300$ GeV, which is 
consistent with the data. In the SDP model, the propagated CR flux can 
be understood as a sum of two components: a harder one due to the 
propagation in the disk and a softer one due to the propagation in the 
halo (see e.g., the discussion in Ref. \cite{2016PhRvD..94l3007F} for 
a simplified two-halo diffusion scenario). 

\subsection{Secondary CRs}

\begin{figure*}[!htb]
\includegraphics[width=0.45\textwidth]{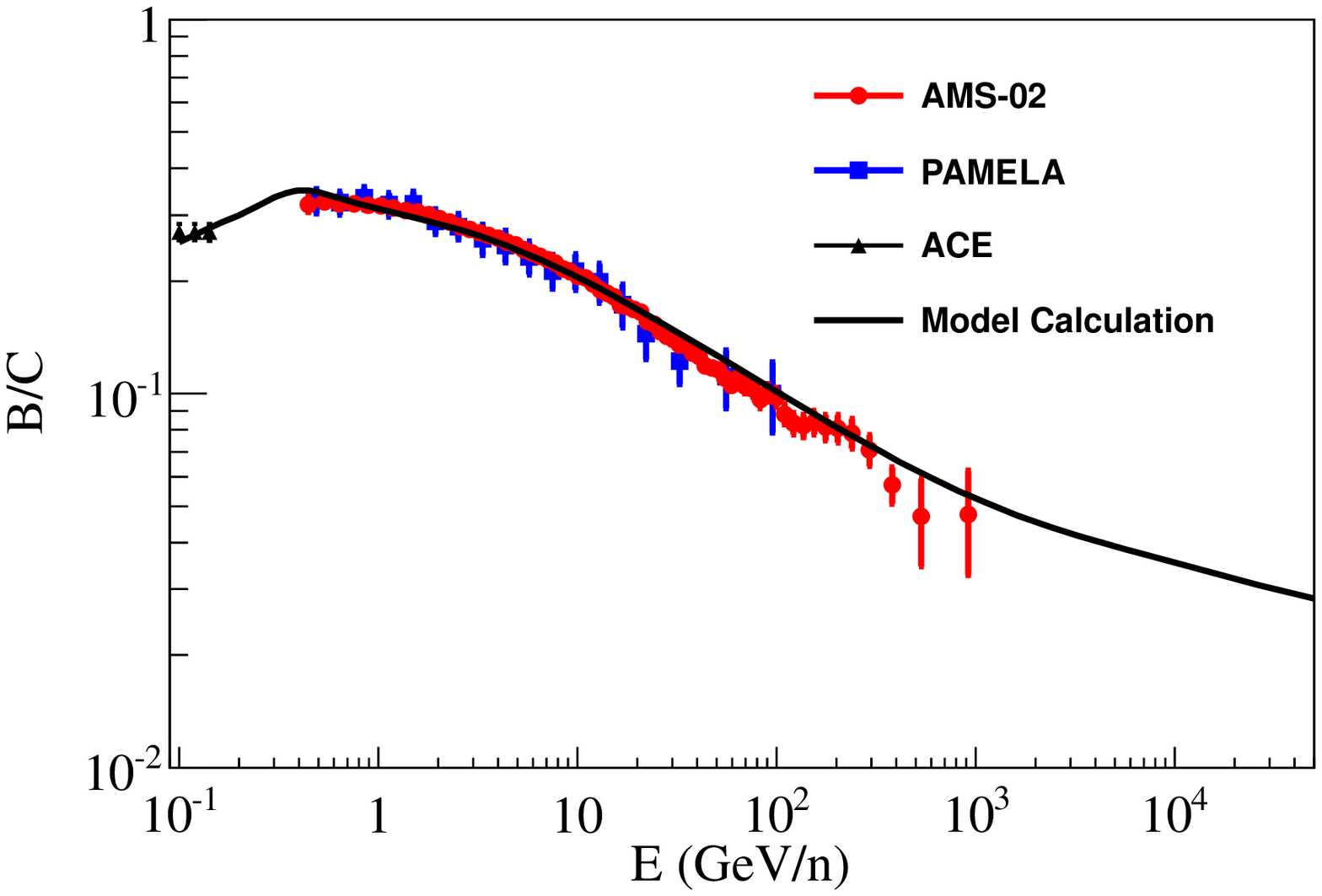}
\includegraphics[width=0.45\textwidth]{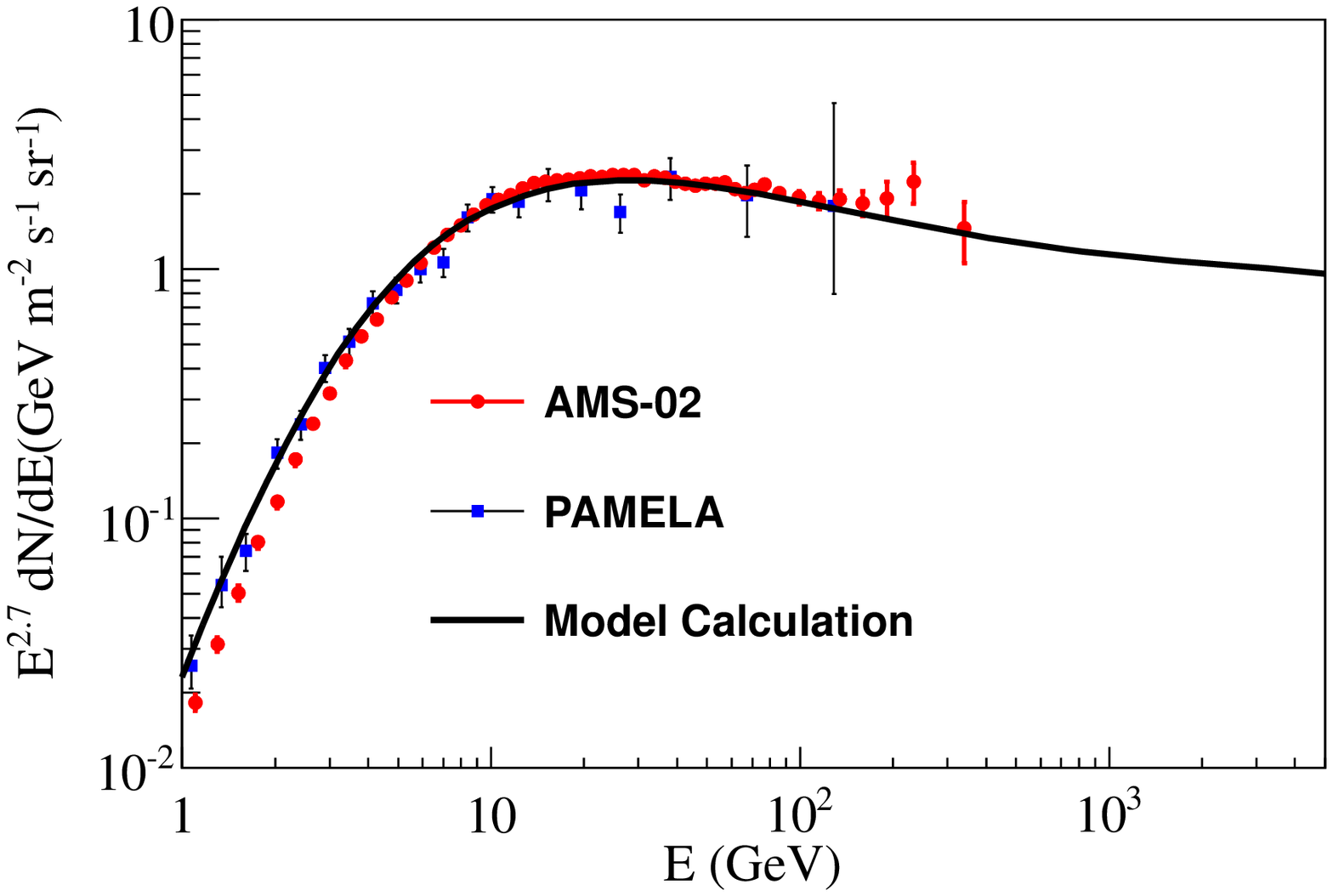}
\caption{Model predictions of the B/C ratio (left) and $\bar{p}$
spectrum (right), compared with the observational data by ACE
\cite{2017PhRvD..95h3007Y}, 
PAMELA \cite{2014ApJ...791...93A,2010PhRvL.105l1101A} and AMS-02 
\cite{2016PhRvL.117w1102A,2016PhRvL.117i1103A}.
\label{fig:BC-pbarp}}
\end{figure*}

Fig. \ref{fig:BC-pbarp} displays the B/C ratio and the $\bar{p}$ spectrum 
predicted by the SDP model. Note that the B/C ratio is in slight tension
with the $\bar{p}/p$ ratio. The AMS-02 data show that, the $\bar{p}/p$ ratio
is almost a constant for rigidities higher than $\sim 60$ GV, while the 
B/C ratio decreases with rigidities following $R^{-1/3}$ 
\cite{2016PhRvL.117w1102A}. The most recent results on the secondary 
Li, Be, and B by the AMS-02 collabration showed that the secondary/primary 
ratios becomes harder by $\sim R^{0.13}$ above 200 GV \cite{AMS02Boron2018}. 
This new result becomes more consistent with the $\bar{p}/p$ ratio, and
seems to support the propagation origin of the spectral hardenings \cite{2017PhRvL.119x1101G}.
Within the uncertainties of the measurements, our model is consistent with 
the data. At high energies, both ratios are expected to harden gradually. 
This is again due to the two-halo propagation nature of particles (secondary 
particles would experience one more time diffusion than primary ones). 
Similar features were also predicted in Ref. \cite{2016ChPhC..40k5001G},
in which a two-component model was proposed to account for behaviors of
secondary particles.
The SDP model can naturally explain the flat behavior of the $\bar{p}/p$ 
ratio above $\sim60$ GeV, without resorting to either astrophysical sources 
\cite{2016PTEP.2016b1E01K,2017PhRvD..96b3006L,2017PhRvD..95l3007C} 
or particle dark matter annihilation \cite{2017PhRvD..95f3021H,
2017JHEP...04..112L,2016arXiv161204001L,2017arXiv170102263F}.

\subsection{Spatial distribution}

\begin{figure*}[!htb]
\includegraphics[width=0.45\textwidth]{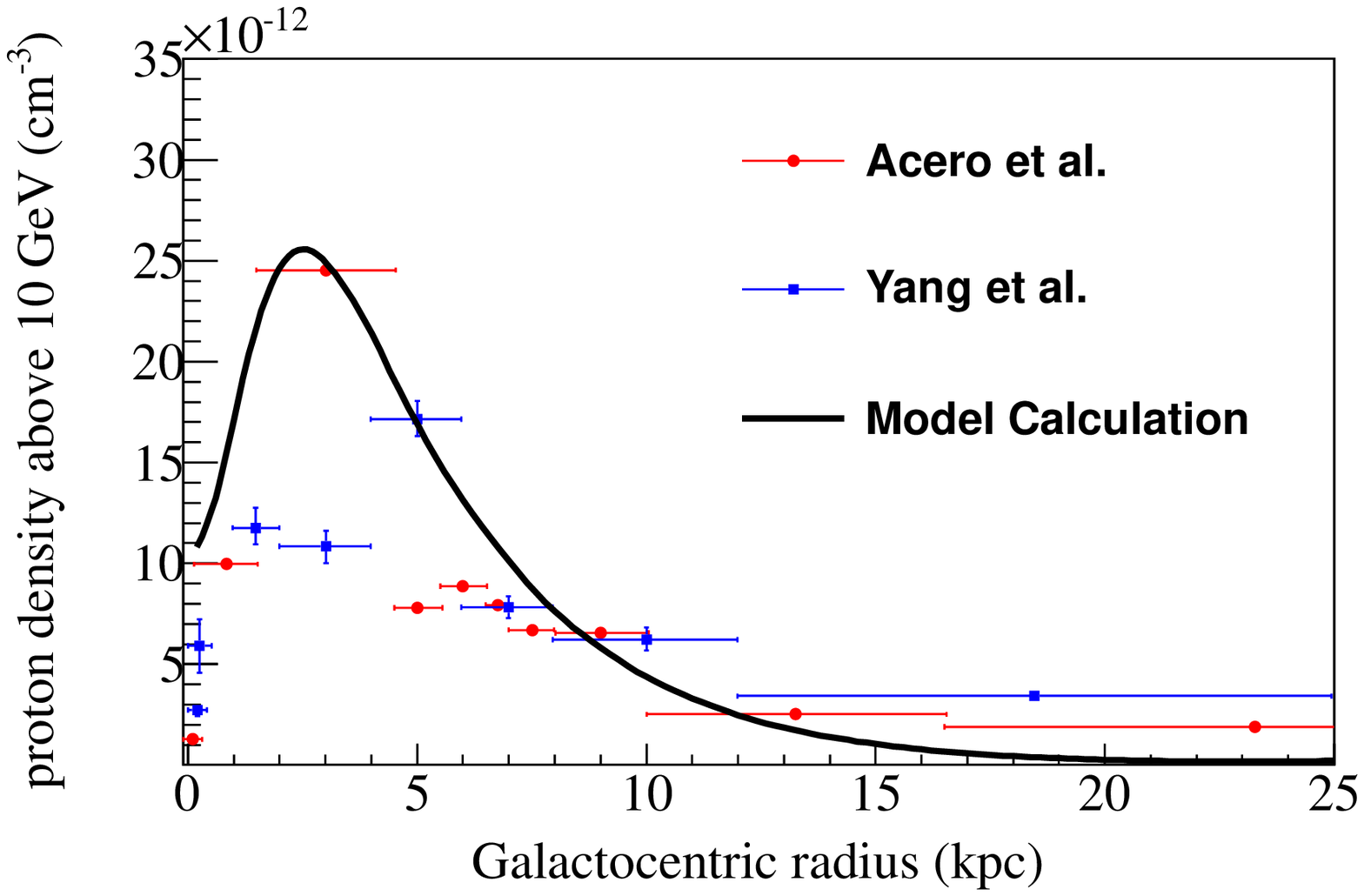}
\includegraphics[width=0.45\textwidth]{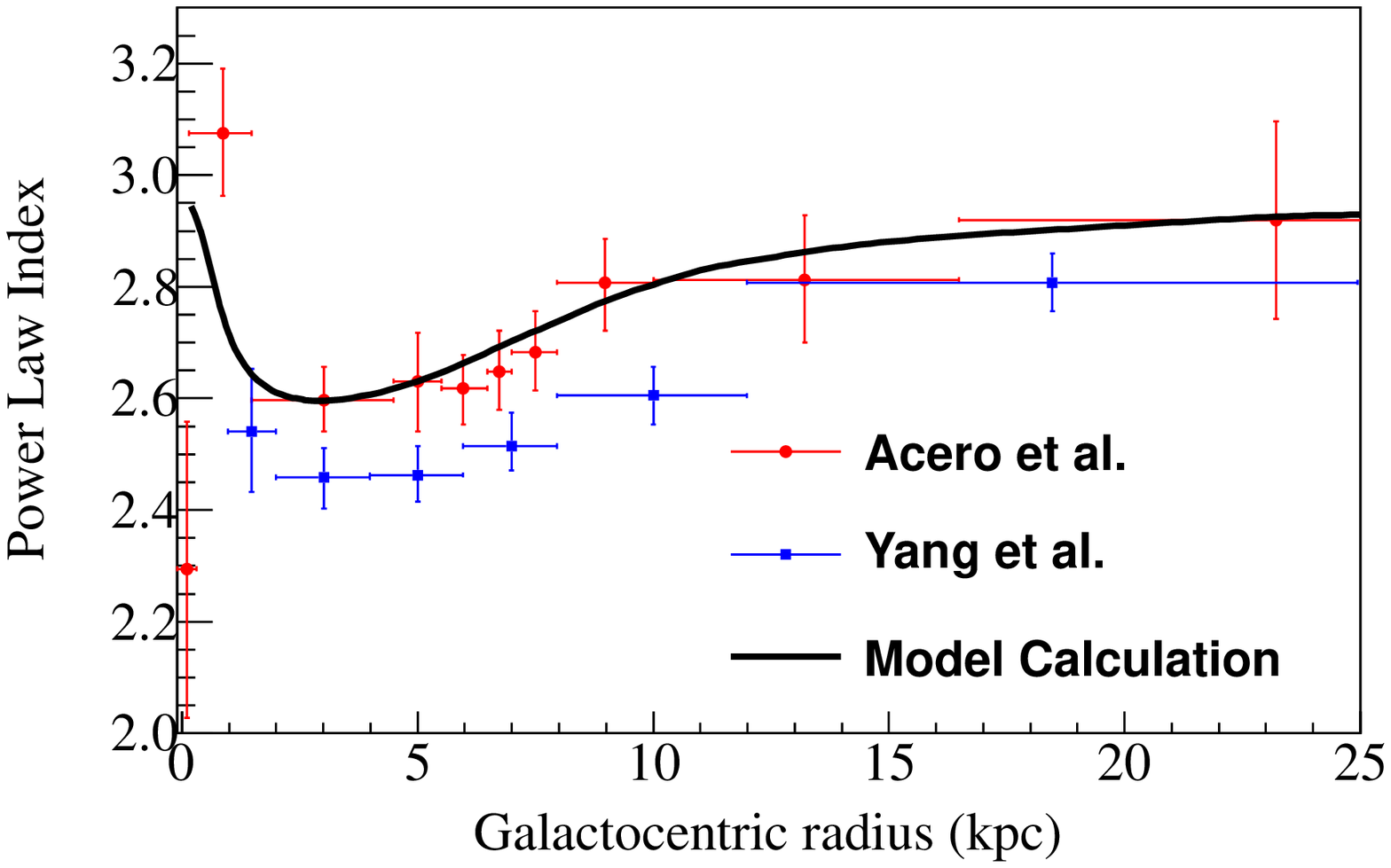}
\caption{Model predictions of the radial distributions of the CR proton 
densities (left) and spectral indices (right), compared with that inferred
from Fermi-LAT $\gamma$-ray data \cite{2016PhRvD..93l3007Y,
2016ApJS..223...26A}.
\label{fig:den-spec}}
\end{figure*}

The spatial distributions of the CR proton densities and spectral indices 
are shown in Fig. \ref{fig:den-spec}, which are broadly consistent with
the results inferred from Fermi-LAT all-sky $\gamma$-ray data
\cite{2016PhRvD..93l3007Y,2016ApJS..223...26A}. Note that these two
analyses differ by a factor of $\sim2$ in the inner Galaxy, probably
due to different gas templates adopted. The quantitative results 
depend on the source parameters, and are thus uncertain to some extent.
Nevertheless, our model correctly reproduce the evolution trends of 
those quantities, especially for the spectral variation. The CR
density reaches a maximum at $\sim3$ kpc, due to the assumed source
distribution of SNRs \cite{1998ApJ...504..761C}.
In the very inner region (Galactic center), the model prediction
is higher than the data. This perhaps requires a non-negligible
advection of CRs in the inner Galaxy, which may result in the formation 
of Fermi bubbles \cite{2010ApJ...724.1044S}. It is also possible 
that the assumed form of the diffusion coefficient of Eq. (3) is not 
precise enough to reveal the diffusion process in the inner Galaxy.
The spectral indicies vary oppositely as the densities. This can be 
understood from the assumed diffusion coefficient. The diffusion 
coefficient is inversely proportional to the source distribution. 
At a few kpc where the source density is the highest, the diffusion 
is the slowest and the rigidity dependence of the diffusion coefficient 
is smallest, therefore the equillibrium CR spectrum is the closest to 
the source spectrum. The spectra become softer for both the inner and 
outer Galaxy regions, where the diffusion is faster.

\subsection{Anisotropies}

\begin{figure}[!htb]
\includegraphics[width=0.48\textwidth]{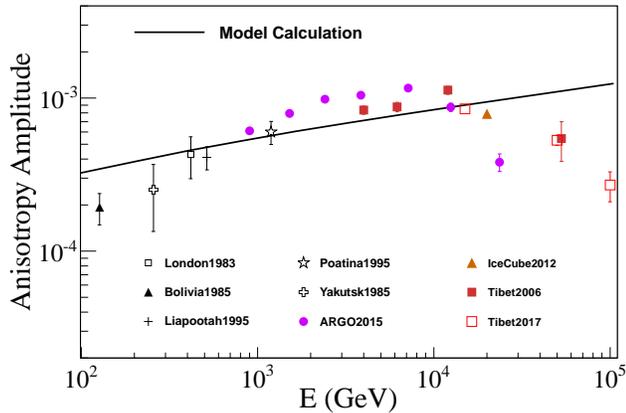}
\caption{The anisotropy of CRs expected from the SPD model, 
compared with the data from underground muon observations: 
London (1983) \cite{1983ICRC....3..383T}, 
Bolivia (1985) \cite{1985P&SS...33.1069S},
Socorro (1985) \cite{1985P&SS...33.1069S}, 
Yakutsk (1985) \cite{1985P&SS...33.1069S},
Liapootah (1995) \cite{1995ICRC....4..639M}, 
Poatina (1995) \cite{1995ICRC....4..635F},
and air shower array experiments: Tibet (2006, 2017) 
\cite{2006Sci...314..439A,2017ApJ...836..153A}, 
IceCube (2012) \cite{2012ApJ...746...33A}, 
and ARGO-YBJ (2015) \cite{2015ApJ...809...90B}.}
\label{fig:aniso}
\end{figure}

The CR anisotropies predicted in the SDP model is lower by nearly an
order of magnitude than that of the conventional diffusion model
\cite{2012ApJ...752L..13T,2016ApJ...819...54G,2017arXiv170107136G}, 
which is consistent with observations below $\sim10$ TeV energies 
\cite{2006Sci...314..439A,2012ApJ...746...33A,2015ApJ...809...90B}, 
as shown in Fig. \ref{fig:aniso}.
We note that, however, the phase evolution of the CR anisotropies
with energies \cite{2017ApJ...836..153A} can not be simply accounted
for by any large scale diffusion model without considering e.g.,
the local source and/or magnetic field effect \cite{2016PhRvL.117o1103A}.

\subsection{Diffuse $\gamma$-ray emission}

\begin{figure*}[!htb]
\includegraphics[width=\textwidth]{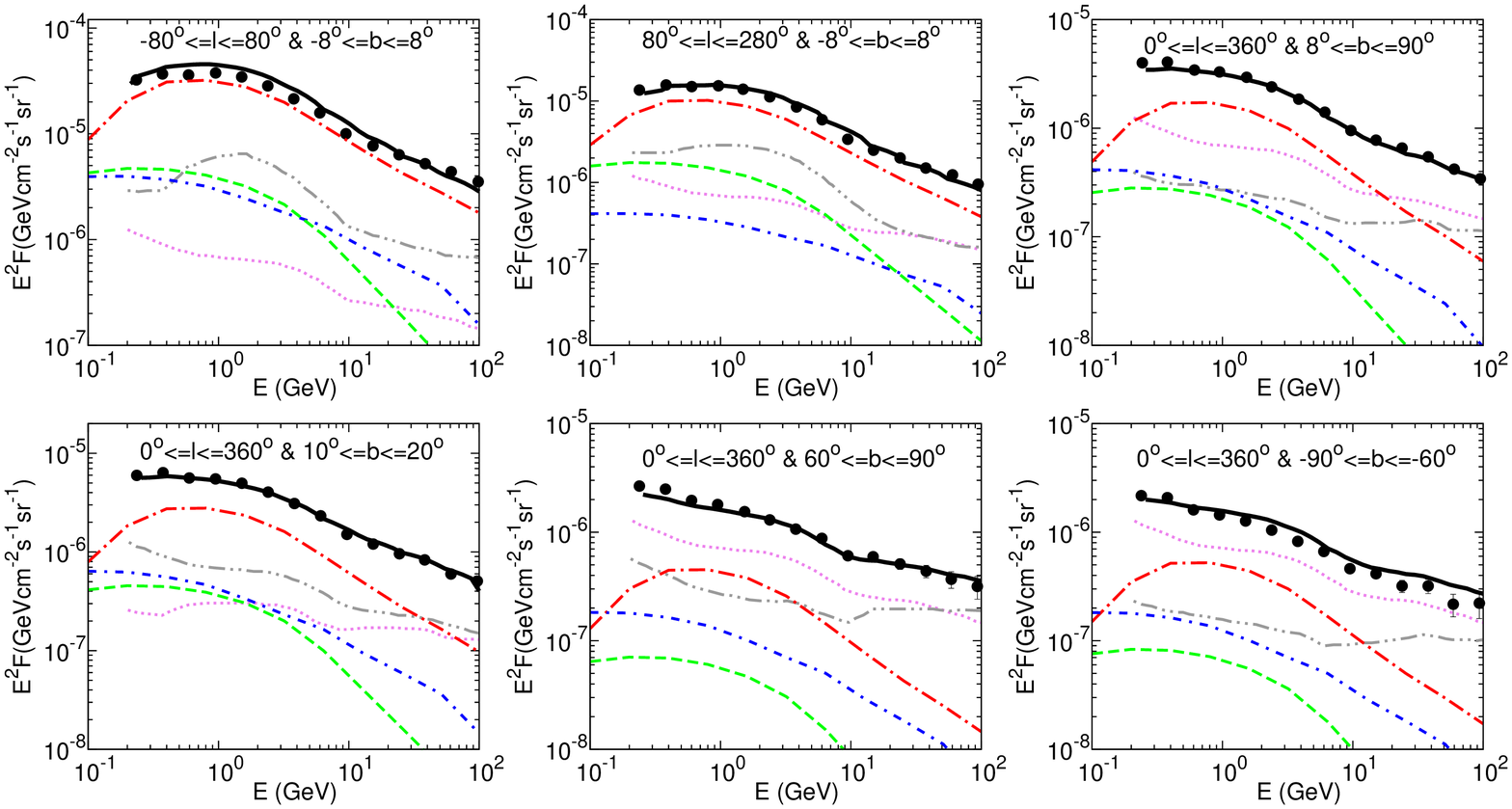}
\caption{Model predictions of diffuse $\gamma$-ray spectra compared with 
observations by Fermi-LAT \cite{2012ApJ...750....3A}. The model calculations 
include the $\pi^0$-decay (red; long dashed-dotted), inverse Compton 
scattering (blue; dashed-dotted), bremsstrahlung (green; dashed) components
of the Galactic diffuse emission, the isotropic background (magenta; dotted), 
and detected sources (gray; dashed-dotted-dotted). The black solid lines
give the total results from the model.
\label{fig:gamma}}
\end{figure*}

The Fermi-LAT observations of diffuse $\gamma$-ray emission are consistent
with the expectation of the conventional CR propagation model at high and
intermediate latitudes, but show excesses in the Galactic plane for energies
above a few GeV \cite{2012ApJ...750....3A}. We show in Fig. \ref{fig:gamma}
the comparison of the $\gamma$-ray spectra in six sky regions between the
SDP model predictions and the data. We find that the Galactic plane excesses
can be well accounted for by the SDP model, due primarily to the spectral
hardening of CRs. The model slightly overproduce $\gamma$-rays in the
inner Galaxy (panel (a)), because of an over-high CR density 
(Fig. \ref{fig:den-spec}). As we have discussed in the subsection III-C, 
an advection of CRs may be present in the Galactic center. 

\section{Discussion}

Many models have been proposed to explain the new observations of CRs,
especially the spectral hardenings. These models can be classified into
several classes. Here we briefly discuss different types of models and
their (potential) performances on different observables.

The modification of the injection spectra of nuclei at source due to
either the intrinsic dispersion of the source properties
\cite{2011PhRvD..84d3002Y} or the non-linear acceleration mechanism
\cite{2013ApJ...763...47P} can make concave particle spectra at
production. In this kind of models, the propagation is assumed to
be the conventional one, and the spectral hardening of CRs is global
in the Milky Way. We may expect that the (high-energy) B/C ratio from
this model is simply follow the inverse of the energy dependence of
the diffusion coefficient, and can thus well fit the data (see for
example, Ref. \cite{2017PhRvD..95h3007Y}). The new data of
secondary Li, Be, and B by AMS-02 favors slightly a break of the
secondary/primary ratio at high energies \cite{AMS02Boron2018}.
Whether this kind of models can be convincingly excluded may need
further studies. The (high-energy) $\bar{p}/p$ ratio should in 
principle decrease with energies. Given the relatively large 
uncertainties of the measurements, the model prediction of $\bar{p}/p$ 
is marginally consistent with the data. Obviously, this model can not 
explain the spatial variation of the CR spectral indices. The gradient
and anisotropy of CRs can not be accounted for either. As for the diffuse
$\gamma$-rays, Ref. \cite{2012ApJ...750....3A} employed the conventional
CR propagation model without considering hardenings of the injection
spectra, and they found excesses at the Galactic plane. To what extent
such mismatches can be solved if the spectral hardenings are included
needs further studies.

\begin{table*}[!htb]
\caption {Summary of different models confronting the observables}
\begin{tabular}{l|c|c|c|c|c|c|c}
\hline \hline
 & Primary   & B/C &  $\bar{p}/p$ & Spatial     & CR  & Diffuse      & Ref. \\
 & hardenings&     &              & distribution& anisotropies& $\gamma$-rays& \\
\hline
Injection       & $\surd$ & \textcircled{}  & \textcircled{} & $\times$ & $\times$ & ? & \cite{2011PhRvD..84d3002Y,2013ApJ...763...47P} \\
Propagation     & $\surd$ & $\surd$ & \textcircled{} & $\times$ & \textcircled{} & ? & \cite{2012ApJ...752...68V} \\
Two components A& $\times$& $\surd$ & $\surd$   & $\times$ & $\times$ & $\surd$ & \cite{2016ChPhC..40k5001G} \\
Two components B& $\surd$ & \textcircled{}  & \textcircled{} & $\times$ & $\times$ & ? & \cite{2015ApJ...803L..15T} \\
Local source    & $\surd$ & $\surd$  & $\surd$   & $\times$ & $\surd$  & $\times$ & \cite{2017PhRvD..96b3006L} \\
Superbubble     & $\surd$ & \textcircled{}  & \textcircled{} & $\times$ & $\times$ & ? & \cite{2011ApJ...729L..13O} \\
SDP             & $\surd$ & $\surd$  & \textcircled{}   & $\surd$  & $\surd$ & \textcircled{} & This work \\
\hline
\hline
\end{tabular}\\
Note: ``$\surd$'' means good agreement, ``$\times$'' means disagreement,
``\textcircled{}'' means marginal agreement, and ``?'' means not clear and
more detailed analysis is needed.
\label{table:summary}
\end{table*}

A second class of models incorporates a pheonomenalogical modification
of the rigidity dependence of the diffusion coefficient, e.g., from
$R^{0.30}$ below 300 GV to $R^{0.15}$ above \cite{2012ApJ...752...68V}.
In a framework that there is a transition of particle interactions
with self-generated turbulence to that with externally generated 
turbulence, Ref. \cite{2012PhRvL.109f1101B} gives such a kind of 
break of the diffusion coefficient. However, quantitatively, they
predicted a rigidity dependence change from $R^{0.7}$ to $R^{0.33}$.
The required break of the diffusion coefficient is currently 
largely empirical.
In this scenario, the B/C and $\bar{p}/p$ ratios would also have breaks
at corresponding rigidities. Since this modification is global in the
Milky Way, the spatial variation of the CR spectral indices can not
be reproduced. The anisotropy of CRs in this model decrease moderately
and can be marginally consistent with the data \cite{2012ApJ...752...68V}.
However, the gradient problem as revealed by Fermi-LAT $\gamma$-rays
should also exist, since it is related to low energy CRs which are the
same for this model and the conventional one. It has been shown that
at intermediate latitudes the diffuse $\gamma$-ray spectrum for this
model is consistent with the Fermi-LAT data \cite{2012ApJ...752...68V}.
However, the consistency with the Galactic plane excesses needs
further studies.

Ref. \cite{2016ChPhC..40k5001G} proposed a two component model (labelled
as ``Two components A'' in Table \ref{table:summary}) to explain the
secondary CR data and the diffuse $\gamma$-rays. In this model, a harder
secondary component is assumed and added to the conventional component.
It has been shown that the $\bar{p}/p$ ratio and diffuse $\gamma$-rays
can be explained. This model also gives a hardening of the B/C ratio at 
high energies, which is consistent with the new data of AMS-02 
\cite{AMS02Boron2018}. However, all the results related to the primary 
CRs, including the spectral hardenings, spatial variations, and gradient
and anisotropies, are not reproduced. 

Ref. \cite{2015ApJ...803L..15T} proposed a model with two types of SNRs
(labelled as ``Two components B'') which have different behaviors of the
secondary production. Secondary particles are not only produced during
the propagation but also around the old SNR population (and get accelerated
meanwhile). The other young SNR population produce harder primary CRs,
but are less efficient in generating secondary particles. This model can
account for the primary spectral hardenings and the featureless B/C ratio
\cite{2015ApJ...803L..15T}. The $\bar{p}/p$ ratio is expected not to be
well reproduced. The spatial variations of the CR spectra are not accounted
for by either, as long as there are no significant differences of the
spatial distributions of these two SNR populations. Since the source
distribution and propagation are similar with the conventional model,
we expect that the gradient and anisotropy problems remain. The diffuse
$\gamma$-ray emission of this model should be similar with the
injection/propagation model.

Some works employed nearby source(s) to account for the either the
primary CR spectral hardenings or the secondary excesses
\cite{2009PhRvD..80f3003F,2012APh....35..449E,2012MNRAS.421.1209T,
2013A&A...555A..48B,2017PhRvD..96b3006L}. In Ref. \cite{2017PhRvD..96b3006L}
it has been shown that adding a nearby source which has effective
interactions with molecular clouds, the primary CR spectra, B/C and
$\bar{p}/p$ ratios, positron and electron fluxes, as well as the
anisotropies can be reasonably accounted for. The contribution of
the nearby source to CRs is mostly local, and hence the CR spatial
variations and diffuse $\gamma$-rays can not be well explained.

Superbubbles have been suggested to be main sources of CRs and are responsible
for the spectral hardenings and He/$p$ ratio \cite{2011ApJ...729L..13O,
2016PhRvD..93h3001O}. This idea is supported by the fact that most of
the Galactic supernovae explode in superbubbles \cite{2005ApJ...628..738H}.
The resulting CR spectral and spatial distributions of this scenario is
similar with that of the ``injection'' model discussed above. Therefore,
the spatial variations and anisotropies may not be well reproduced.

We summarize the comparison of different models with different observables
in Table \ref{table:summary}. Since the spatial variations of CR spectra
and the diffuse $\gamma$-rays require changes of the global properties
of CR injection and/or propagation, all models except the SDP model can
reasonably give such results. Furthermore, most of models face the
difficulty to be consistent with the CR gradient and anisotropies.
In the local source model, the anisotropies can be small only when finely
tuned model parameters are adopted (source location is the anti-Galactic
center direction) to cancel the anisotropies from the diffusion of
Galactic CRs. The SDP model can easily decrease the gradient and
anisotropies to be consistent with the data.

\section{Summary}

In this work we suggest an SDP model of Galactic CRs to account for 
the new observational features of CRs and diffuse $\gamma$-rays. 
The SDP model introduces an anti-correlation between the diffusion 
properties and the source distribution of CRs. The physical origin
of this anti-correlation is natural: the turbulent magnetic field
which regulates the diffusion of particles is correlated with the
matter distribution. This simple extension of the conventional uniform 
diffusion model explains the primary spectral hardenings, 
secondary-to-primary ratios, spatial variations of CR intensities 
and spectra inferred from Fermi-LAT diffuse $\gamma$-rays, CR anisotropies, 
and the Galactic plane excesses of diffuse $\gamma$-rays. Compared with 
other proposals of modifications of the conventional CR origin and/or 
propagation model, the SDP model can explain the most of observations, 
with little tuning of the model parameters.

\acknowledgments
We thank Siming Liu and Ruizhi Yang for helpful discussion.
This work is supported by the National Key Research and Development 
Program of China (No. 2016YFA0400200), the National Natural Science 
Foundation of China (Nos. 11635011, 11761141001, 11663006, 11722328), 
and the 100 Talents program of Chinese Academy of Sciences.

\bibliographystyle{apsrev}
\bibliography{grad}

\end{document}